\documentstyle [prl,aps,preprint]{revtex}

\def\Y{$\hbox{Y}\hbox{Ba}_{2}\hbox{Cu}_{3}\hbox{O}_{7-\delta}$}

\def\sqr#1#2{{\vcenter{\hrule height.#2pt
    \hbox{\vrule width.#2pt height#1pt \kern#1pt
      \vrule width.#2pt}
    \hrule height.#2pt}}}

\begin{document}
\draft 
\title{Possibility of Macroscopic Resonant Tunneling near the
Superconductor-Insulator Transition \\ in \Y ~Thin Films}
\author{S. Tanda, K. Kagawa, T. Maeno, T. Nakayama and K. Yamaya} 
\address{Department of Applied Physics, Hokkaido University, Sapporo 060-8628, Japan
} 
\author{A. Ohi} 
\address{Fuji Electric Corp. Research and Development, Ltd., 
Yokosuka, Kanagawa, 240-01, Japan}
\author{N. Hatakenaka} 
\address{NTT Basic Research Laboratories, Atsugi, Kanagawa, 243-01, Japan} 

\date{Received 2 September,1997}
\maketitle
\begin{abstract} 
Experimental results of the $I-V$ characteristics near
the superconductor-insulator transition observed for disorder-tuned 
$\hbox{Y}\hbox{Ba}_{2}\hbox{Cu}_{3}\hbox{O}_{7-\delta}$ 
thin films are presented. 
The $I-V$ characteristics 
exhibit new quasiperiodic structures as a function of the current. 
The current interval, the number of the $dI/dV$ peaks, and 
the magnetic field dependence of the peaks are consistent with the 
theoretical predictions of the {\it resonant tunneling} of a phase particle 
in a tilted-cosine potential for a single Josephson junction 
with small capacitance.

\end{abstract}

\baselineskip 0.92cm

Quautum phase transition in two-dimensional (2D) systems has been the subject 
of much theoretical and experimental works for the sake of its 
abundant physical implications. 
In paticular, the superconductor-insulator (S-I) transition 
related to the quantum unit of resistance $h/4e^{2} \sim 6.45$ k$\Omega$ 
has been studied for various materials 
\cite{Goldman,Haviland,Heb,Tanda,Sei,Kuma,Kagawa,Deu}. Among them, 
cuprate superconductors with 2D $CuO_2$ sheets 
are suitable to study this phase transition 
because disorder can easily be controlled by oxygen deficiency.
These systems also enable us to investigate the macroscopic quantum 
tunneling (MQT) \cite{Caldeira}, 
in connection with the applicability of quantum mechanics 
on a macroscopic scale, 
because of the spontaneous formations of Josephson networks with small 
capacitances due either to the presence of crystal grains 
or to oxygen contamination in \Y\ thin films, or both. 
In such a small Josephson junction, 
the phase difference $\theta$ across the junction cannot be considered 
as a classical variable; it must be treated as an operator that does not 
commute with the Cooper-pair-number operator $\hat{n}$; 
$[\hat{\theta}, \hat{n}]=i$. 
Under current biases, the junction can be represented as a {\it quantum} 
particle moving in a one-dimensional tilted cosine potential. 
The escape from the metastable state by quantum 
tunneling is thus possible.  

Another quantum-mechanical process, resonant tunneling, is possible 
in Josephson systems. 
When the charging energy becomes much larger than that of MQT, 
the quantized energy levels formed in the Josephson potential 
become important for tunneling. 
Under the current bias, the alignment of the levels between the adjacent wells 
rules the tunnel phenomena. 
This is known as macroscopic resonant tunneling (MRT) \cite{Hatakenaka}. 
Giordano and Schuler have observed 
the quasiperiodic structures of current-voltage characteristics 
due to the resonant features of the tunneling processes, {\it i.e.}, 
the enhancement and supression of $2\pi$ quantum phase slippages  
in one-dimensinal superconducting $Pb$ wires \cite{Giordano}. 
Although the system is similar to Josephson systems, 
theory remains ambiguous, 
especially regarding the periodicity of the potential.
This Letter demonstrates MRT phenomena near 
the S-I transition in \Y\  thin films under magnetic fields. 
First, we will show that the \Y\  thin films can be regarded as a series of 
Josephson junctions. 
Next, we present the I-V characteristics of the system. 
We found that the $I-V$ curves near the S-I transition 
exhibit quasiperiodic structures as a function of the current similar to 
the Giordano and Schuler's observation. 
The structures strongly depend on the magnetic field. 
Last, we compare our results to theoretical predictions of
the $resonant$ $tunneling$ of a phase particle  
in a small Josephson junction.

    The \Y\ thin films were deposited on MgO (100) substrates by 
rf-magnetron sputtering using a \Y\ disk target. 
The sputtering gas was a mixture of Ar (50\%) and $\hbox{O}_{2}$ (50\%) 
with total pressure of 0.5 Pa. 
The substrate temperature was kept at 650~C  during the deposition process. 
The temperature dependence of resistivities $\rho(T)$ was measured 
with evaporated gold electrods 
by the standard four terminal method, 
and the applied current was typically 1 to 10~mA. 
The differential resistances were measured by a lock-in amplifier 
with the frequency of 364~Hz. 
Figure 1 shows the sheet resistances as a function 
of temperature for six samples (A-F). 
The average mean of the sheet resistance per CuO$_2$ layer 
in \Y\ thin films and the parameter representing the intensity 
were obtained from the 
relation $R_{\Box} = \rho /d$, where $d$ (= 5.9~\AA) is 
the lattice spacing between CuO$_2$ layers. 
Although the disordered parameters in these films were not systematically 
tuned, the observed normal-state sheet resistances 
above $T_{c}$ are different enough 
that the degree of disorder of these samples is widely distributed.  
The critical sheet resistance has a value of about 
10~k$\Omega$ close to the value $h/4e^{2} \sim 6.45~$k$\Omega$. 
The curves B-E show local coherence. 
Each resistance exhibits a pronounced drop within the 60 to 70~K range, 
while these samples did not show global coherence 
and had finite resistance. 
In addition, a decrease in temperature led to an increase in resistance. 

In Fig.~2 we plotted the normalized conductivity $ \sigma / \sigma_0 $ 
for sample F as a function of $(T-T_{c}) / T_c$ on a logarithmic scale, 
where $\sigma_0$  is the normal-state conductivity. 
The data clearly show the power law in the 
regime  $(T-T_{c}) / T_{c} \le 1$, and we have the relation  
$\sigma \propto (T-T_c)^{-s}$ with $s =1.36\pm 0.03$.
This estimated value is in good agreement with the 
theoretical prediction  $s=1.30$ \cite{Straley}. 
The agreement indicates that the superconductor 
in sample F can be regarded as a 2D 
superconducting percolation system, which is the fundamental model for 
strongly disordered 2D systems.

As the temperature is reduced, the superconducting network 
gradually extends.  
At the critical temperature $T_c$, the superconducting path 
extends from top to bottom and from left to right of the 
sample, and the whole sample becomes globally coherent.  
The superconducting percolation network is comprised of links - 
one-dimentional chains - and blobs - dense regions with more than one 
connection between two points. 
If one link is cut at $T_c$,  
the global coherence is broken and this link constitutes a backborn 
structure. 
Compared with globally coherent sample F, 
the superconducting path is reduced considerably 
in the large-disordered samples B-E.  
Therefore, links are regarded as Josephson junctions 
in series and blobs are regarded as superconductors with zero resistivity. 
Locally coherent samples B-E with large disorders are 
considered ultrasmall Josephson junctions in series, 
$I-V$ characteristics are determined by the smallest Josephson junction. 
The multi-junction system is nonetheless preferable for observing 
the behavior of a {\it single} Josephson junction 
because the junctions in series provide high resistance 
and supress the effect of the electromagnetic 
environments \cite{Delsing}. 

We measured the $I-V$ characteristics near the S-I transition 
under various magnetic fields (Fig.~3). 
We found that the $dV/dI-I$ curves 
exhibit quasiperiodic structures as a function of the current. 
These structures are similar to 
the Giordano and Schuler's observation, 
which strongly depends on the magnetic field. 
Let us explain these new structures from the standpoint of MRT. 
First we calculated $E_{\rm J}$ using the Ambegaokar-Baratoff equation 
\cite{Ambegaokar}, 
$E_{\rm J}= \pi \hbar \Delta/4e^{2}R_{N}$,  at zero temperature, 
where $\Delta$ is the gap energy and $R_T$ is the tunnel resistance. 
The $R_T$ is roughly given by the normal resistance of the system, 
$R_T\simeq R_N$, which is different from $R_{\Box}$. 
In sample E, the resistance is $R_{N}=120 ~\Omega$,  
and the gap energy $\Delta=20$~meV.  
Thus, the Josephson coupling energy $E_{\rm J}$ is 
about $8.6 \times 10^{-20} ~{\rm J}$. 
In contrast to the Josephson coupling energy, it is difficut to 
estimate the charging energy $E_c$ of the junction because 
no techniques have been established 
for estimating the capacitance of the junctions in random media. 
The junction capacitance is usually evaluated from the junction geometry 
or the Coulomb gap in I-V characteristics. 
We applied the Coulomb gap technique to our system, but we could not 
obtained the junction capacitance 
because sample E does not satisfy the Coulomb blockade 
conditions, $R_N > h/e^2\sim 25~$k$\Omega$.  
Thus we were obliged to infer the capacitance 
from the value of the other samples. 
Fortunately, the capacitance of sample B, with large resistance 
$R_N=40~$k$\Omega$, can be obtained. 
Figure 4 shows the I-V characteristics of sample B.
The curve clearly shows the Coulomb gap due to single electron tunneling, 
indicating the formation of small junctions in the sample and strongly 
supporting our Josephson model for \Y ~thin films.
The capacitance of the junction in sample B, 
estimated from the Coulomb gap $(e/2C=71~$mV), 
is $C=1.1\times 10^{-18}$~F. 
The charging energy $E_c$ is about $1.1 \times 10^{-20} ~{\rm J}$. 
The capacitance of sample E can be evaluated through the ratio of resistance 
between the junctions on different samples, since both capacitance and 
resistance are characterized by the junction's cross-section 
\cite{Capacitance}. 
The charging energy of sample E 
is on the order of $4.2 \times 10^{-21} ~{\rm J}$.

Now let us explain our experimental results in terms of 
macroscopic resonant tunneling. 
First, we check whether our experimental conditions were such that MRT 
occurred. 
MRT requires quantized energy levels in the Josephson potential. 
The number of quantum levels ($n$) in the potential is obtained as  
the ratio of the height of the Josephson potential $2E_{\rm J}$ to 
the level spacing, roughly  $\hbar \omega_{\rm J}$, 
in the harmonic approximation of the potential; 
\begin{equation}
\ n \sim  \frac{2E_{\rm J}}{\hbar \omega_{\rm J}}
     =\sqrt{\frac{E_{\rm J}}{2E_c}}.                   
\end{equation}
Here $\omega_{\rm J}$ is the Josephson plasma frequency. 
The number of levels $n$ is at least more than 2 for MRT to occur. 
Moreover, the temperature must be held low enough to 
prevent masking by the thermal fluctuations,{\it i.e.}, k$_B T \ll 
\hbar \omega_{\rm J}$. In sample E,  all of the above conditions are satisfied;
 
$n \simeq 3 $ and $T=4.2$~K.  
In fact, we observed that periodic structures are broadened
as the temperature increases, and disappear at $T=55$~K. 
This implies that the origin of the structures stems from 
the quantum-mechanical effects. 

Next, we have compared our results to the MRT predictions. 
MRT occurs when the energy levels in adjacent wells are coincident. 
These coincidences occur at special values of bias currents 
which reflect the quantized energy levels. 
First, we checked the periodicity of the oscillations in $dV/dI-I$ curves. 
The current intervals can be determined by setting the level spacing equal to 
the energy gain from the battery, $\hbar \omega_{\rm J}= \hbar 
\Delta I/2e$; thus, 
\begin{equation}
\Delta I=\frac{2e}{h}\sqrt{8E_{c}E_{\rm J}},                  
\end{equation}
where we use the relation $\omega_{\rm J}=\sqrt{8E_cE_{\rm J}}/\hbar$. 
The current interval is $\Delta I \sim 2.4 ~\mu$A. 
This is close to the experimental value $\Delta I \sim 6~\mu$A, 
in spite of the rough estimation of the capacitance. 
Second, 
the number of the observed peaks is consistent with the number of quantum 
levels in the potential calculated above. 
Third, the upper peak position is also consistent with the MRT 
predictions; as the current bias increases, the tunneling mechanism changes 
from MRT to usual MQT. 
We can determine the threshold current by 
considering the potential's arrangement \cite{Hatakenaka}. 
In our case, $E_c/E_{\rm J}=0.07$ and the threshold current is 
therefore about $0.17I_c=44~\mu$A, where 
$I_c$ is the Josephson critical current. 
The current of third peak, $25~\mu$A, is less than the threshold current.  
Furthermore, the magnetic field dependence of the $dV/dI-I$ curves 
is also consistent with the theoretical predictions; 
as the magnetic field was increased, 
the number of peaks reduced from three to two 
and the intensity of the peaks gradually weakened. 
Hence, 
the strong magnetic field reduces the Josephson coupling energy $E_{\rm J}$ 
and prevents the occurrence of the MRT. 
Finally, the structure disappeared at $0.25$~Tesla. 
Just as with the temperature, this reflects quantum-mechanical effects. 
Thus, we have concluded that the new quasiperiodic 
structures in I-V characteristics are due to the MRT.

In summary, we have found that the $I-V$ characteristics near 
the universal critical sheet resistance exhibit 
new quasiperiodic structures as a function of the current in  \Y\  thin films.
The periodic structures are in good agreement with theoretical predictions 
of the MRT in small Josephson junctions.

We would like to 
thank Prof. N. Giordano of Purdue University 
for his invaluable discussion. 
We also thank the Kurata Foundation and the Iketani Science 
Foundation. 

\newpage
{\noindent 

{Fig.~1 The temperature dependence of the resistivity of 
disorder-tuned \Y thin films. Film thickness were 300\AA (A,B), 1000\AA (C,D,E,
F).\\}}

{Fig.~2 The temperature dependence of the conductivity of sample F. The solid l
ines show the power law with s=1.36 $\pm$ 0.03. \\}

{Fig.~3 $dV/dI-I$ curves of sample E. The temperature was 4.2K. \\}

{Fig.~4 $I-V$ characteristics of sample B. The threshold voltage was 71mV.}


\begin{references}
\bibitem{Capacitance}
In general, both the capacitance and the resistance of the junction 
are the function of both tunnel distance $d$ and cross-section $S$, 
and can be described by $C \sim S/d$ and $R_N\sim e^{-d}/S$, respectively. 
Our 2D Josephson networks are made by introducing 
oxygen deficiencies in the $CuO_2$ planes of \Y. 
Since the deficiency has the same effect as cutting the bonding,  
only the cross-section of the junction is changed 
while tunneling distance seems to be constant.
\end{references}
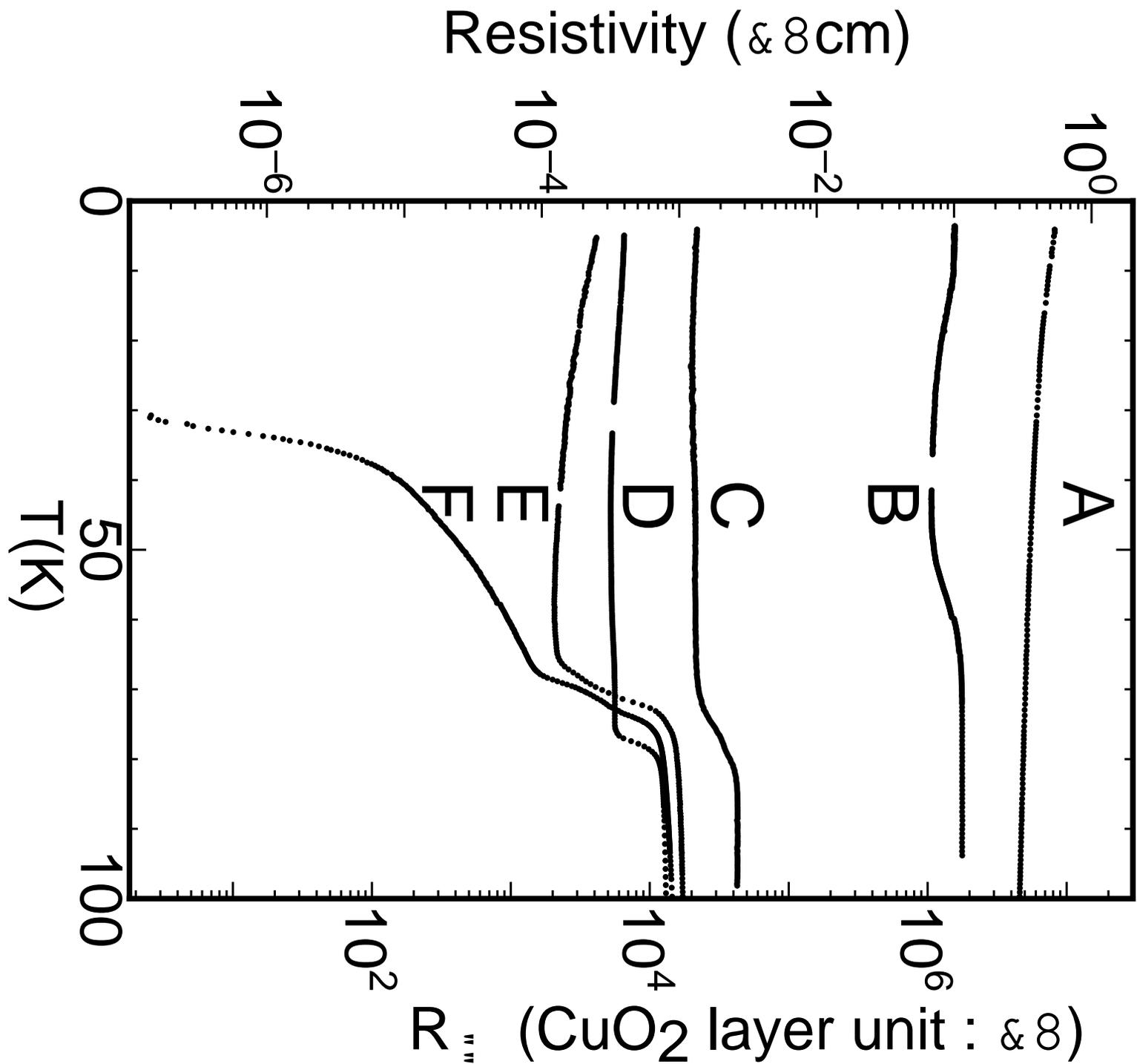
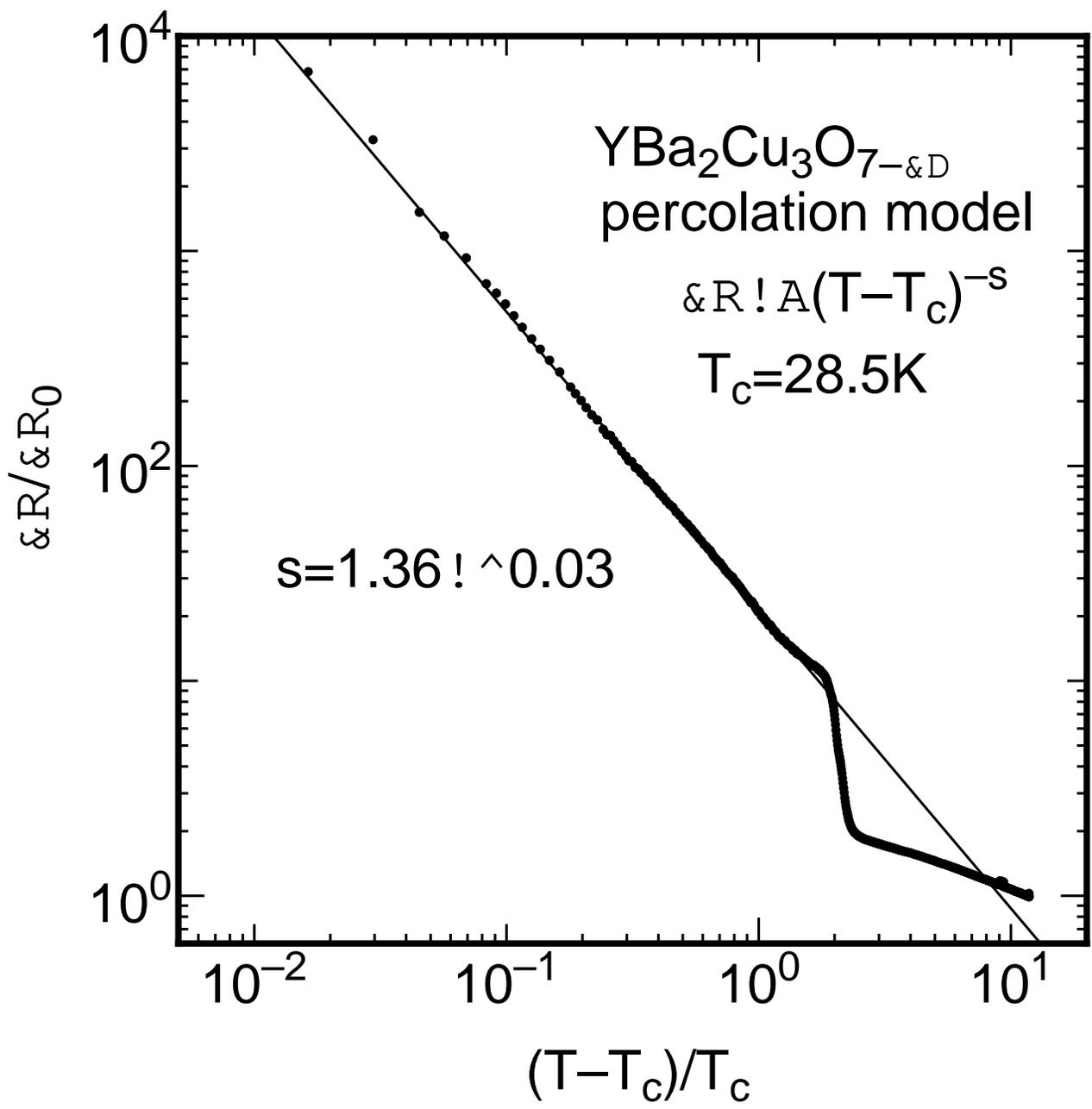
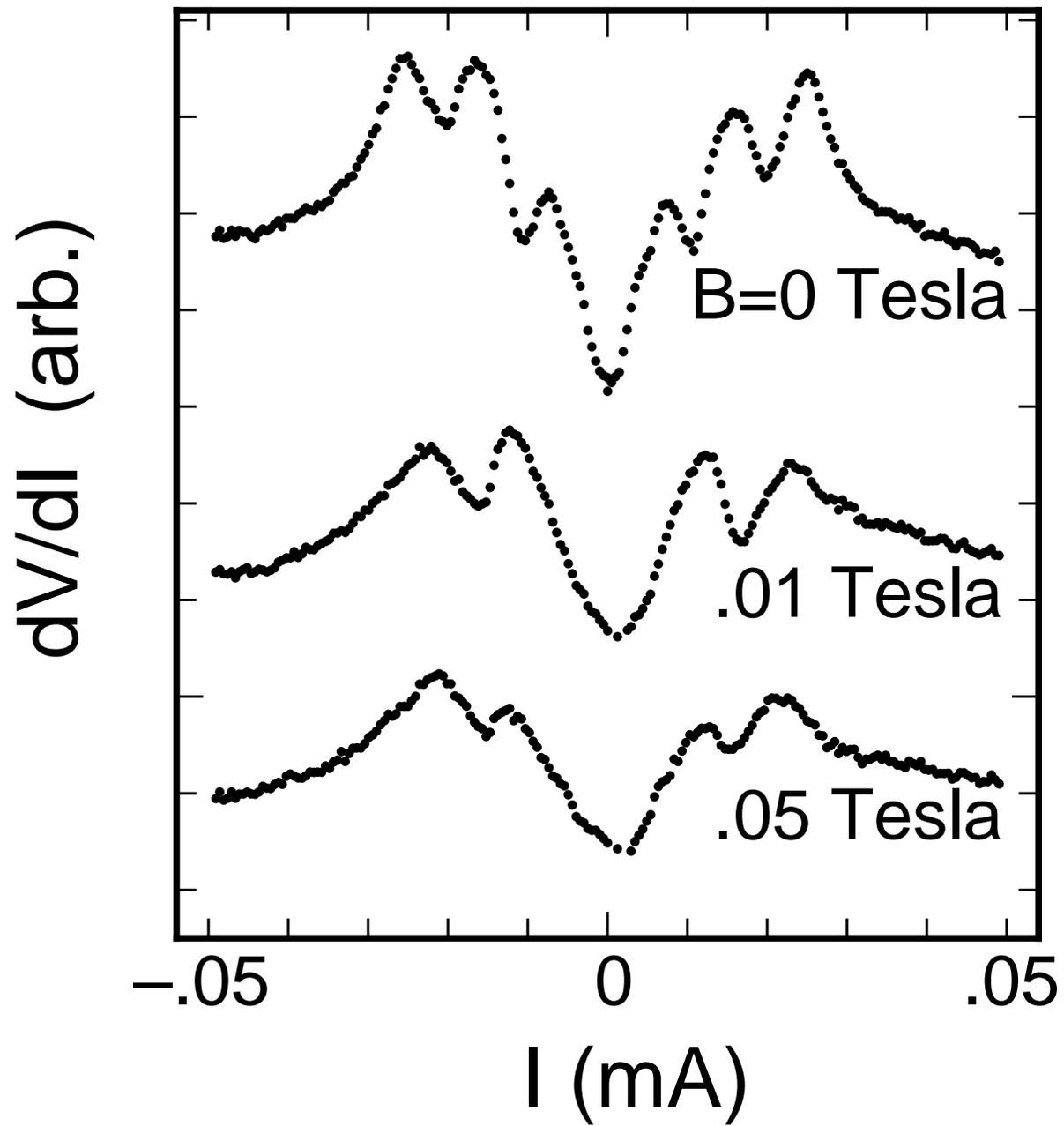
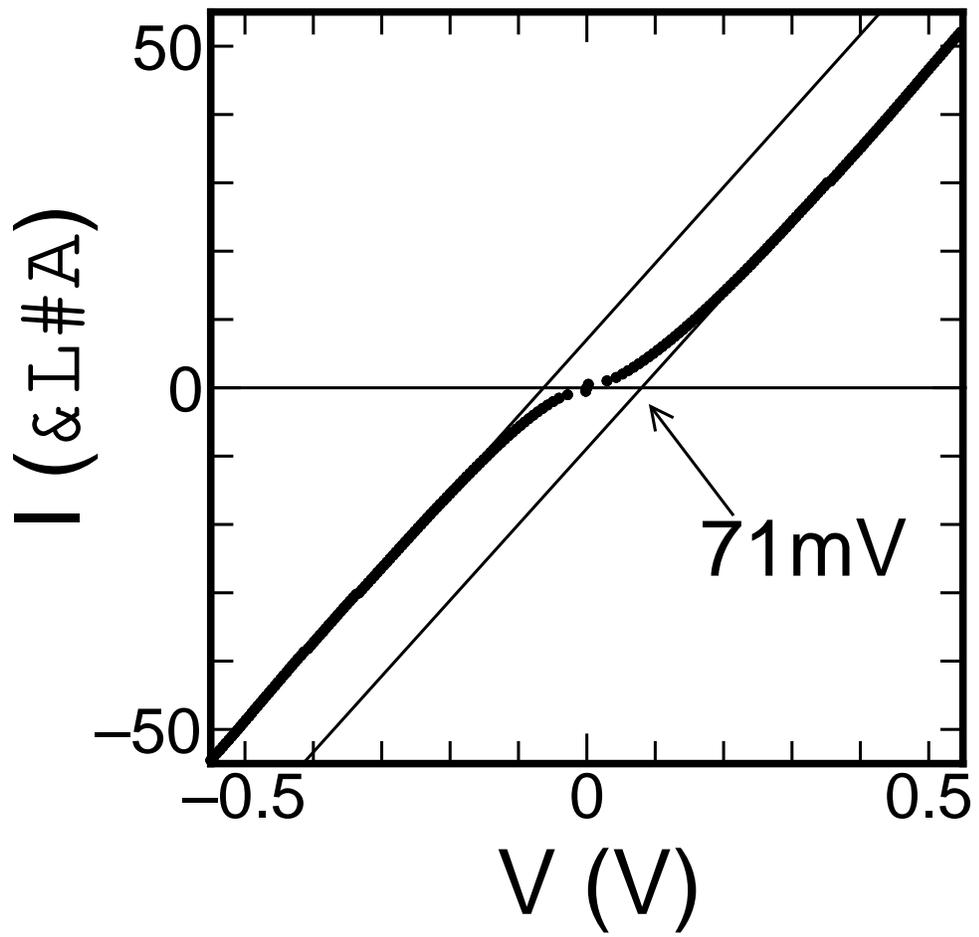
\end{document}